\documentclass{article}

\usepackage{nomencl}                    
\usepackage{float}                      
\usepackage[numbers]{natbib}            
\usepackage{graphicx}					
\usepackage{fancyhdr}                   
\usepackage{url}                        
\usepackage{relsize}                    
\usepackage{booktabs}                   
\usepackage{caption}
\usepackage{subfig} 
\usepackage{mathrsfs}                   
\usepackage{array}
\usepackage{amsmath}
\usepackage{paralist}
\usepackage{pdflscape}
\usepackage{amssymb,amsthm,fancyhdr,mathabx}

\begin{document}

\title{Developing Synthetic Spectroscopy Noise and Chemometric Database for Computational Classification}


\author{ { Nicholas J. Napoli}
      \thanks{Please request code of the the noise generator or additional documentation} \\
	Dept. of Electrical and Computer Engineering\\
	University of Florida\\
	Gainesville, FL 32611 \\
	\texttt{n.napoli@ufl.edu} \\
}

\maketitle

\begin{abstract}
There has been little to no work in the area of spectroscopy noise in order to create data sets for analytical algorithms to be challenged on the ability to separate chemicals. We present a framework on how to build off of a sparse about of experimental data in order to expand your chemometric database and create realistic instrumentation noise. The combination of various interactions of chemicals combined with various random permutations of spectroscopy noises enables researches to better capture and model the multitude types of signals and variations that can be present within an experimental reading.
\end{abstract}


%

\section{Introduction}


The unavailability of open source spectroscopy data and the difficulty in direct acquisition of spectroscopy data via experiments are significant hurdles in the development of effective detection algorithms associated with such data in chemometrics. To address this issue, we sought to develop a method to generate a synthetic spectroscopy data bank which attempts to faithfully capture the operationally relevant features of fluorescent spectra. We wanted this method to possess the ability to account for the variations arising from different excitation wavelengths, multiple combinations of chemical mixtures, different apparatus and effects of user error.

\section{Data from PhotoChemCAD}\label{Sec:ChemicalDatabase}
To generate such a synthetic database, we started with a sparse amount of reputable data from PhotoChemCAD \cite{Du, Dixon}. To achieve our objective with this limited set of experimental data, we made several assumptions:

\begin{compactitem}
  \item There is no chemical quenching.
  \item The spectral resolution is accurately depicted through interpolation and decimation.
  \item All path lengths are $1\,cm$.
  \item Accurate depiction of the molar absorptivity is projected by scaling the absorption spectrum by cited molar extinction coefficients.
  \item Beer-Lambert's law is obeyed.
  \item Concentrations are below $1\mu M$ to minimized inner filter effects to achieve additive absorption and emission spectra.
\end{compactitem}
From PhotoChemCAD, we chose multiple chemicals within the same solvent, Toluene. The analytes selected were from two chemical classes: oligopyrrole and polycyclic aromatic hydrocarbons. These two criteria were selected to obtain data that are similar in chemical composition. The resulting strong spectral similarities create a ``non-trivial" data set for classification and quantification purposes. The chemicals that were used in the database are listed in Table~\ref{tab:ChemicalListA} and \ref{tab:ChemicalListB}.

The data from PhotoChemCAD provide neither the concentration of the analytes nor the emission characteristics at different excitation wavelengths. In our data set, we acquire these two quantities directly from published results for spectral modeling see Table~\ref{tab:ChemicalListA} and \ref{tab:ChemicalListB}. Epsilon, $\varepsilon$, enables us to account for different concentrations of the chemical; and the quantum yield, $\Phi_F$, enables us to calculate different emissions spectra at various excitation wavelengths. With these quantities accounted for, we are able to generate a more realistic data set.

\begin{table*}[h]
\center
  \begin{tabular}{|l|c|m{2.55cm}|m{2cm}|c|}
    \hline
    \hfil\textbf{Chemical}
      & \textbf{Solvent}
      & \textbf{Epsilon ($\varepsilon$) in $cm^{-1}/M$ at $\lambda_{Ex}$}
      & \textbf{Quantum Yield ($\Phi_F$)}
      & \textbf{Cited} \\
    \hline\hline
    5,10-Diaryl Chlorin & Toluene & $89,100$ at $414\,nm$ & $0.260$ & \cite{Taniguchi} \\
    \hline
    5,10-Diaryl Mg-oxoChlorin & Toluene & $191,000$ at $408\,nm$  & $0.100$ & \cite{Taniguchi} \\
    \hline
    5,10-Diaryl oxoChlorin & Toluene & $174,000$ at $414\,nm$  & $0.130$ & \cite{Taniguchi} \\
    \hline
    5,10-Diaryl Zn-Chlorin & Toluene & $186,000$ at $412\,nm$ & $0.083$ & \cite{Strachana, Taniguchi} \\
    \hline
    5,10-Diaryl Zn-oxoChlorin & Toluene & $209,000$ at $408\,nm$  & $0.040$ & \cite{Taniguchi} \\
    \hline
    Bis(5-mesityldiprinato)zinc & Toluene & $115,000$ at $487\,nm$ & $0.360$ & \cite{Sazanovich} \\
    \hline
    Bis(5-phenyldiprinato)zinc & Toluene & $115,000$ at $485\,nm$  & $0.006$ & \cite{Sazanovich} \\
    \hline
    Magnesium Octaethylporphyrin & Toluene & $408,300$ at $410\,nm$ & $0.150$ & \cite{Zass, Yang} \\
    \hline
    Magnesium Tetramesityporphyrin & Toluene & $446,700$ at $426.5\,nm$ & $0.170$ & \cite{Lindsey, Yang} \\
    \hline
    Magnesium Tetraphenylporphyrin & Toluene & $22,000$ at $564\, nm$ & $0.150$ & \cite{Miller, Strachanb} \\
    \hline
  \end{tabular}
  \caption{Chemicals in the Database: Oligopyrrole.}
  \label{tab:ChemicalListA}
\end{table*}

\begin{table*}[h]
\center
  \begin{tabular}{|l|c|m{2.55cm}|m{2cm}|c|}
    \hline
    \hfil\textbf{Chemical}
      & \textbf{Solvent}
      & \textbf{Epsilon ($\varepsilon$) in $cm^{-1}/M$ at $\lambda_{Ex}$}
      & \textbf{Quantum Yield ($\Phi_F$)}
      & \textbf{Cited} \\
    \hline  \hline
    Perylene-diimide & Toluene & $44,000$ at $490\,nm$ & $0.97$ & \cite{Prathapan} \\
    \hline
    Perylene-Monoimide & Toluene & $32,000$ at $511\,nm$ & $0.86$ & \cite{Tomizaki} \\
    \hline
    Perylene-Monoimide(OR)3 & Toluene & $32,000$ at $479\,nm$ & $0.91$ & \cite{Tomizaki} \\
    \hline
    Perylene-Monoimide (OR) & Toluene & $40,000$ at $507\,nm$ & $0.82$ & \cite{Tomizaki} \\
    \hline
  \end{tabular}
  \caption{Database Chemicals: Polycyclic Aromatic Hydrocarbons.}
  \label{tab:ChemicalListB}
\end{table*}

\subsection{Preprocessing of Spectral Data }

The spectral data from PhotoChemCAD is an agglomeration of various sources, where acquisition parameters are distinctive from each other. The spectral data from the selected chemicals are different in their wavelength range and optical resolution. To ensure proper manipulation among different chemical spectral vectors, the length and indexing of the vectors are required to be equivalent. The optical sampling of the spectral vectors obtained were either $0.25\,nm$, $0.5\,nm$, or $1.0\,nm$; hence, the vectors were interpolated or decimated to get values corresponding to $0.5\,nm$. Each vector was padded with elements of value $1.0\cdot 10^{-20}$ to provide a uniform wavelength range throughout the entire database of selected chemicals. The value $1.0 \cdot 10^{-20}$ is used to avoid absolute zero errors and to circumvent subsequent complications with vector and matrix manipulations.

\subsection{Generating Spectra Corresponding to Various Concentrations}

A spectrometer takes measurements of light absorption, producing a unique spectral absorption signature. When taking measurements, the concentration and pathlength are held constant. The epsilon value, which is a function of the excitation wavelength, is an intrinsic property of the measured chemical that defines the spectral waveform characteristics. The selected data only provides the absorbance, giving no insight into the concentration or pathlength. We are therefore unable to distinguish the epsilon values due to the unknown collection parameters from the various sources provided by PhotoChemCAD. The measured absorbance of the sample is proportional to the number of absorbing molecules from the incident light of the spectrometer and it is essential that the absorbance value is corrected for a meaningful comparison \cite{Reusch}. This correction for absorption is referred to as \emph{molar absorptivity} or \emph{molar extinction coefficients}, which serves to compare spectra and evaluate the relative strength of the absorbance. In order perform a proper comparison between spectra, we scale the spectral vector with respect to epsilon at its appropriate listed excitation wavelength from Table~\ref{tab:ChemicalListA} and \ref{tab:ChemicalListB}.

Consider a measured absorption data vector from Table~\ref{tab:ChemicalListA}:
\begin{equation}
  \textbf{D}_{Ab_j}
    =\begin{bmatrix}
       d_{Ab_{j1}} & d_{Ab_{j2}} & \cdots & d_{Ab_{jN}}
     \end{bmatrix}^T,
\end{equation}
where $d_{Ab_{jk}}\in[0, A], \forall k\in\overline{1,N}$, $A$ is an arbitrary positive number, and $j$ is a specific chemical. These optical absorption measurements were scaled to coincide with cited molar extinction coefficients (i.e., epsilon) at the corresponding wavelength from Table~\ref{tab:ChemicalListA} and \ref{tab:ChemicalListB} via the following equation:
\begin{equation}
\label{eq:MolarAb}
  \textbf S_{Ab_j}
    =\textbf{D}_{Ab_j}
     \frac{\varepsilon_{\lambda_{Ex} j}}{D_{Ab_j\lambda{_{Ex}}}}
     C,
\end{equation}
where $\varepsilon_{\lambda_{Ex} j}$ is the molar extinction
%
%
coefficient of chemical $j$ from Table~\ref{tab:ChemicalListA} and \ref{tab:ChemicalListB}, $C$ is the concentration, and $D_{Ab_j\lambda_{Ex}}$ is the element in $\textbf{D}_{Ab_j}$ that is associated with epsilon at the specific excitation wavelength of $\lambda_{Ex}$. With the spectral vector properly scaled, we assume that each  spectral element depicts its appropriate molar extinction coefficient for all wavelengths. We can now apply Beer-Lambert's law to expand the database by altering the concentration while we hold constant the pathlength at $1\,cm$.

\subsection{Expanding the Data Set Via Quantum Yield}\label{Sec:quantumyield}

When a molecule is excited to a higher quantum state of a particle and it transitions to a lower state, the molecule emits a photon. The more the molecule absorbs energy the higher potential for it to elicit more photons. The amount of fluorescence emission is a function of the amount of light absorbed by a molecule. This function is known as the \emph{quantum yield of the fluorescence, $\Phi_F$.} It is defined as the number of photons emitted over the total number of photons absorbed \cite{Szabo}. Based on the intrinsic nature of the molecule and its absorption properties, specific wavelengths are more prone to be absorbed than others. It is apparent that the excitation wavelength affects the total intensity of the absorption and concurrently affects the total emission intensity. By dynamically changing the excitation wavelength, we generate various emission spectra accounting for the effect of quantum yield.

Consider a measured fluorescence emission spectrum vector from Table~\ref{tab:ChemicalListA} and \ref{tab:ChemicalListB}:
\begin{equation}
  \textbf{D}_{Em_j}
    =\begin{bmatrix}
       d_{em_{j1}} & d_{em_{j2}} & \cdots & d_{em_{jN}}
     \end{bmatrix}^T,
\end{equation}
where $D_{em_{jk}}\in[0, A], \forall k\in\overline{1,N}$ , $A$ is an arbitrary positive number, and $j$ is a specific chemical. In order to generate further data and for a realistic simulation of different excitation wavelengths, we consider $I_{o\lambda}$ as the intensity of incident light to excite the sample. We are able to quantify the summed intensity of the emission fluorescence, $S_{Em}$, as
\begin{equation}\label{eq:QYield}
  \textbf S_{Em}
    \propto
     \textbf N_{Em_j} S_{Ab_j\lambda_{\varepsilon}}
     \Phi_{F_j}I_{o\lambda },
\end{equation}
where $S_{Ab_j\lambda_{\varepsilon}}$ is the element in $\textbf{S}_{Ab_j}$ that is associated with epsilon at $\lambda_{Ex}$ and $\textbf N_{Em_j}$ is the normalized vector of $\textbf{S}_{Em_j}$, i.e.,
\begin{equation}
  \textbf N_{Em_j}
    =\frac{\textbf{D}_{Em_j}}
     {\displaystyle\sum_{k=1}^N \textbf{D}_{Em_j}}.
\end{equation}

\begin{figure}[h]
  \centering
  \includegraphics [width=.9\linewidth]{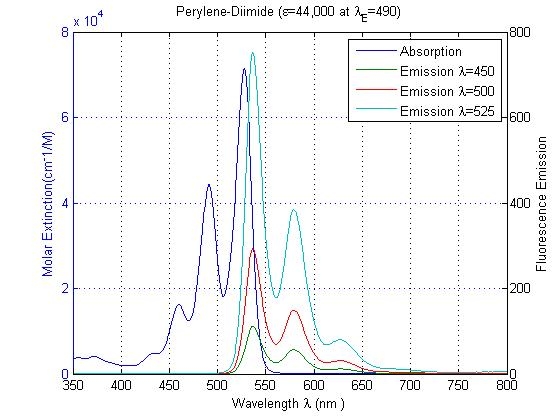}
  \caption{Emissions as a Function of $\lambda$ and $\Phi_F$.}   \label{fig:ExcitEmiss}
\end{figure}

Note that the summed fluorescence emission is dependent on the incident light intensity, the absorbance magnitude at a particular $\lambda_{Ex}$, and $\Phi_F$ \cite{Szabo}. As shown in Figure~\ref{fig:ExcitEmiss}, different $\lambda_{Ex}$\,s obtained from the absorption spectrum elicit different energy contributions to the spectral topology
of the emission signal.

\subsection{Generating Spectra Corresponding to Combinations of Chemicals}

Contingent on our assumption that the absorption and emission spectra are additive, we expand our database to include linear combinations of chemicals. This assumption is valid only when Beer-Lambert's law is obeyed and the inner filter effect is minimized, thus allowing us to define the linear combination process at the wavelength of interest as
\begin{equation}
  \label{eq:BeerAdditive}
  A^{\lambda_1}_{x+y}
    =A^{\lambda_1}_x+A^{\lambda_1}_y
    =\epsilon^{\lambda_1}_{x}bC_x+\epsilon^{\lambda_1}_{y}bC_y.
\end{equation}

One way to introduce an amount of different chemicals is by sampling without replacement and without ordering. This can be accomplished via~\ref{eq:CombinA} and expressed as a binomial coefficient, where $k$ is the number of analytes that are chosen from a set of $n$ total number of analytes \cite{Garcia}:
\begin{equation} \label{eq:CombinA}
  C^{n}_{k}
    =\frac{n(n-1)\ldots (n-k+1)}{k!}
    =\frac{n!}{k!(n-k)!}
    =\left(
       \begin{matrix}
         n \\
         k
       \end{matrix}
     \right)
\end{equation}
However, our intention is to evaluate these analytes over different combinations of $k$. A visual example of four different chemicals is shown in Table~\ref{tab:CombinA}.
\begin{table}[H]
  \center
  \begin{tabular}{|c|c|c|c|}
    \hline
    $k=1$ & $k=2$ & $k=3$ & $k=4$ \\
    \hline\hline
    Chem(1) & Chem(1,4) & Chem(1,2,3) & Chem(1,2,3,4) \\
    Chem(2) & Chem(4,2) & Chem(1,2,4) & \\
    Chem(3) & Chem(4,3) & Chem(1,3,4) & \\
    Chem(4) & Chem(3,1) & Chem(2,3,4) & \\
    & Chem(3,2) && \\
    & Chem(2,1) && \\
    \hline
    Total = 4 & Total = 6 & Total = 4 & Total = 1 \\
    \hline
  \end{tabular}
  \caption{Evaluating Chemical Permutations.}
  \label{tab:CombinA}
\end{table}
Hence, we need to define equation~\ref{eq:CombinA} over a sum of all $k$:
\begin{equation}
  \label{eq:CombinB}
  \sum^{k=n}_{k=1}\frac{n!}{k!(n-k)!}
    =\sum^{k=n}_{k=1}
     \left(
       \begin{matrix}
         n \\
         k
       \end{matrix}
     \right).
\end{equation}

It is apparent that, as we evaluate various $n$ number of sets using equation~\ref{eq:CombinB}, we can get $2^n-1$ different chemical combinations (the $-1$ is due to the fact that we are not considering the scenario where there is no chemical present within the database). These combinational permutations can also be later leveraged by utilizing probability detection paradigms and matching algorithms to the classification of spectroscopy signal~\cite{Napoli_DS_EKG,NapoliMap}, 

For software implementation purpose, we use a binary representation for chemical presence/absence within the sample.  This is illustrated by amending Table~\ref{tab:CombinA} with the appropriate coded binary representation, where each bit represents chemical presence/absence ($1$=Chemical Present and $0$=Chemical Not Present).
\begin{table*}[H]
  \center
  \begin{tabular}{|c|c|c|c|}
    \hline
    k=1 & k=2 & k=3 & k=4 \\
    \hline\hline
    Chem(1)=[0001] & Chem(1,4)=[1001] & Chem(1,2,3)=[0111]
      & Chem(1,2,3,4)=[1111] \\
    Chem(2)=[0010] & Chem(4,2)=[1010] & Chem(1,2,4)=[1011]
      & \\
    Chem(3)=[0100] & Chem(4,3)=[1100] & Chem(1,3,4)=[1101]
      & \\
    Chem(4)=[1000] & Chem(3,1)=[0101] & Chem(2,3,4)=[1110]
      & \\
    & Chem(3,2)=[0110] & & \\
    & Chem(2,1)=[0011] & & \\
    \hline
  \end{tabular}
  \label{tab:BinaryComb}
  \caption{Binary Coding of Chemicals.}
\end{table*}

\section{Errors in Spectroscopy Measurements}
Spectroscopy signals are also affected by electronic noise, stray light, light scattering, wavelength accuracy, resolution, stability, baseline flatness, effects of  sampling geometry, and user error \cite{Lakowicz}. While it is not realistic to accommodate all these types of errors, we now discuss how several additional sources of error are introduced into our synthetic spectroscopy data set.

\subsection{User Error}

\begin{figure}[H]
  \centering
  \includegraphics [width=.9\linewidth]{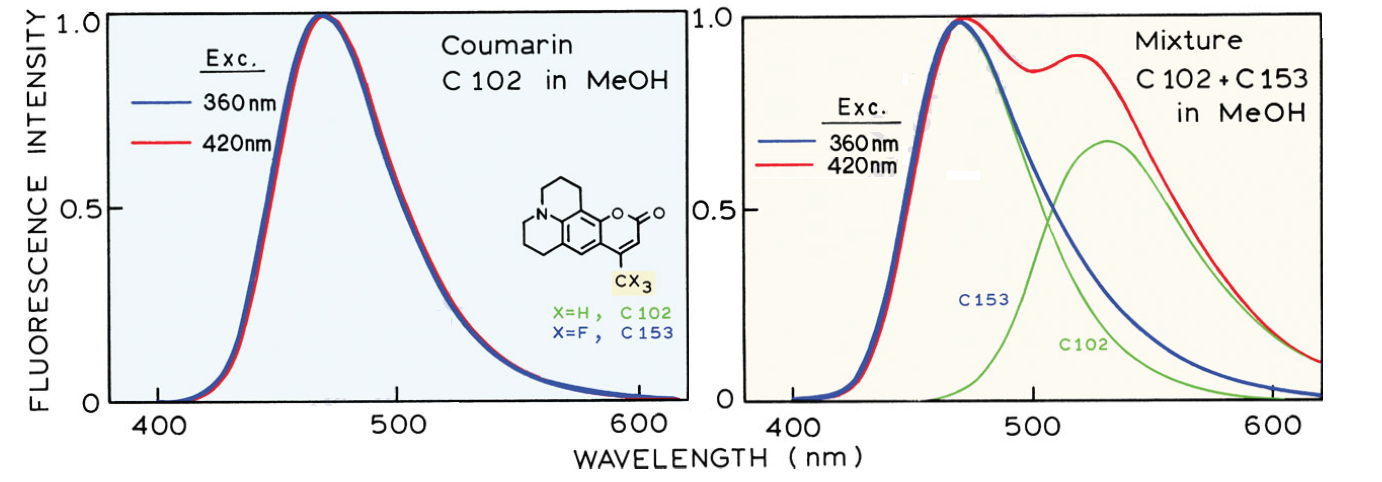}
  \caption{Emission Spectra of $C102$ and a Mixture of $C102$ with the Fluorescent Impurity $C153$ \cite{Lakowicz}.}
  \label{fig:ExciteImpurity}
\end{figure}

Use error is quite common in spectroscopy data. The most common user error involves a lack of concentration, usually associated with pipetting chemical dilutions at low concentrations. This error can be further exacerbated if the molar absorptivity is high. Applying Beer-Lambert's law to this error, one would expect changes in peak height and overall spectral area. Another user error involves fluorescent contamination of the measured sample, or when the detected light is contaminated by Rayleigh or Raman scatter. This is also contingent on the particle size of the analyte, which is a function of the variance within the measured spectrum \cite{Barnes}. Figure~\ref{fig:ExciteImpurity} shows how the emission remains the same with different excitation wavelengths in a pure sample, and how a contaminate alters the emission spectrum topology at $420\,nm$  \cite{Lakowicz}.

\subsection{Stray Light}

\begin{figure}[h]
  \centering
  \includegraphics [width=.9\linewidth]{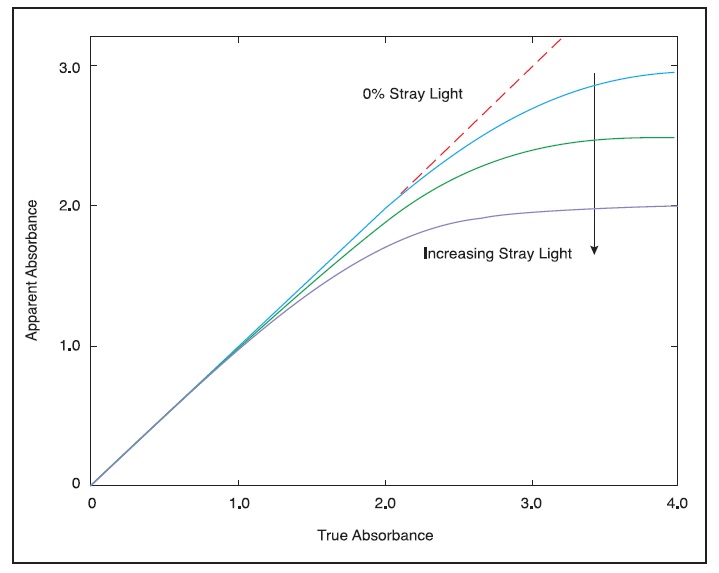}
  \caption{Apparent Absorbance vs True Absorbance with Increasing Stray Light \cite{Allen}.}
  \label{fig:StrayLight}
\end{figure}

Stray light is the measured light of any wavelength reaching the detector that is not associated with the bandwidth of the selected wavelength \cite{Allen}. Stray light manifests itself as an apparent deviation in Beer-Lambert law. The effects of stray light is a decrease in absorbance and a reduction of the perceived projected linearity of the absorbance. This can be described by equation~\ref{eq:TransmittanceAbs}, where $I$ is the transmitted light, $I_s$ is the stray light, and $I_o$ is the incident light:
\begin{equation}
  \label{eq:TransmittanceAbs}
  Absorbance
    =-\log
     \left(
       \frac{I+I_s}{I_o +I_s}
     \right).
\end{equation}
Figure~\ref{fig:StrayLight} shows the effect of stray light on the absorbance \cite{Allen}.

\subsection{Wavelength Accuracy}
\begin{figure}[h]
  \centering
  \includegraphics[width=.9\linewidth]{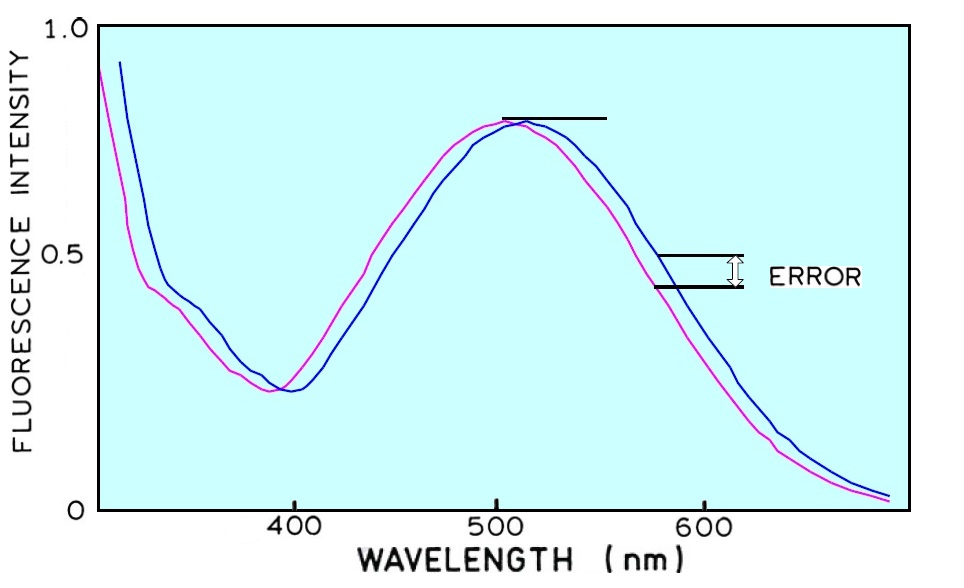}
  \caption{Wavelength Accuracy \cite{Lakowicz}.}
  \label{fig:WaveAccuracy}
\end{figure}

Wavelength accuracy is the inability to preserve the wavelength scaling at the detector or emitter. This scaling error introduces a shift in the measured wavelength. This causes our perception of the true $\lambda_{max}$ to be inaccurate \cite{Lakowicz, Allen}. See Figure~\ref{fig:WaveAccuracy}.

\subsection{Self-Absorption}
\begin{figure}[h]
  \centering
  \includegraphics[width=.9\linewidth]{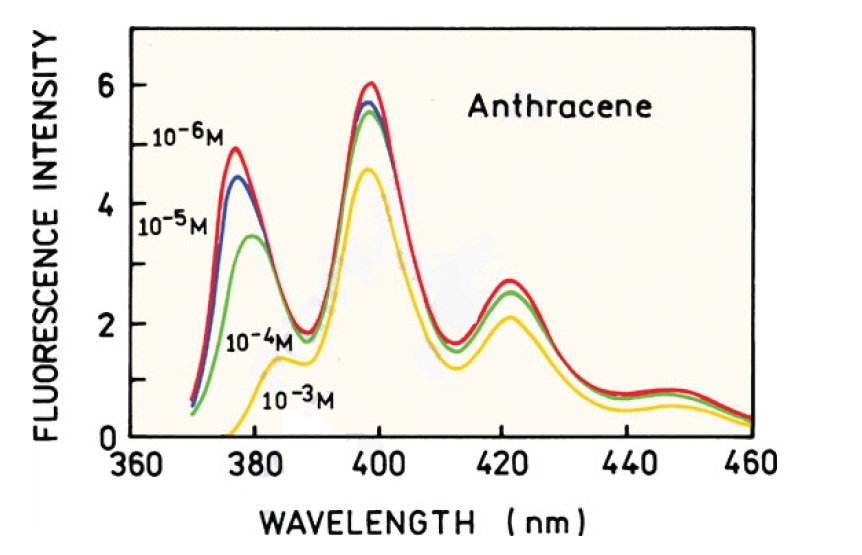}
  \caption{Effects of Self-Absorption.}
  \label{fig:Selfabsorbation}
\end{figure}

Self-absorption depends upon the geometric arrangement observing the fluorescence and high optical densities, which can cause intensity distortion within specific wavelength ranges. As can be seen in Figure~\ref{fig:WaveAccuracy}, the error causes a shifting of the spectrum. Figure~\ref{fig:Selfabsorbation} is an example of a right-angle observation, where short wavelength emissions are attenuated by the analyte Anthracence's absorbance at the shorter wavelengths \cite{Lakowicz}.

\subsection{Overview of Spectroscopy Errors}
From the discussion above, one notices that a large proportion of the errors associated with spectroscopy measurements cause the nominal spectrum to be perturbed in a ``smoother" manner. Other types of errors that generate higher frequency perturbations in the spectroscopy measurements are typically modeled as additive white gaussian noise (AWGN). For the current purpose, our intention is to model the errors that have a higher potential to elicit a misclassification within the database. This is what we undertake in the following sections.

\section{Insertion of Spectroscopy Measurement Perturbations}
\label{Sec:SyntheticCorrelatedNoiseA}
The introduction of perturbations to the database allows for the creation of more realistic test samples in our prototype data set. This enables us to explore how signal degradation can affect classification performance of our algorithm.

\subsection{Windowing}
We employ different windowing functions as the basic strategy to alter the spectral emission and absorption vectors.
The \textit{windowing foundation} is only the basis of defining, where the types of window functions will be implemented in the spectra. These locations for each type of windowing function shares a relationship with the spectral peaks of emission and absorption vectors. In later sections, the windowing functions are modified to provided either compression or dilation to the spectra within a specified design range.

Consider the following spectral data vector (corresponding to emission or absorption):
\begin{equation}
  \textbf{S}
    =\begin{bmatrix}
       S_1 & S_2 & \cdots & S_N
     \end{bmatrix}^T,
\end{equation}
where $S_k\in[0,A],\,\forall k\in\overline{1,N}$, and $A$ is an arbitrary real positive constant. The peaks of this vector were examined to adaptively create perturbations in the vicinity of these peaks.
This was done to create a unique correlated noise (Dilation or Compression) to individual spectral vector. This correlated noise is dependent on $\textbf S$'s spectral peak ``shape", the windowing function type ($t$), and the given window size $L_w$, where $L_w\ll N$. The peaks were sought by implementing the Matlab function \texttt{findpeaks}, where we set a minimum window distance $L_w$ between peaks. The \texttt{findpeaks} function yields a vector $\textbf P$, where the maximum peak location for the processed spectra are represent by the elements of $\textbf P$:
\begin{equation}
  \textbf{P}
    =[k_{1}, \hspace{5 mm}   k_{2}, \hspace{2 mm}
      \cdots k_{K}]^T,
\end{equation}
where $k_p\in[1,N], \forall p \in \overline{1,K},$ and $K\leq N/L_w$. Given $L_w$ and $N$, we determine the number of windows that will be designed by taking the integer quotient $\lfloor N/L_w \rfloor=C$, where $C$ is the number of windows to be designed. This yields $C+1$, window segments. We determine the type of windowing function that will be employed to each window segment, $i$, based on our \emph{peak indictor} vector, $\textbf{I}_p$. The peak indictor vector informs us at what window segment a peaks occur by taking the integer quotient plus a unit ones vector $\textbf U_K$, where $K$ is the length.
\begin{equation}
\textbf{I}_p=\lfloor \textbf P/L_w \rfloor+ \textbf U_{K}
\end{equation}
We use vector $\textbf I_p$ to determine, $t$, the type of windowing function to implement for $\textbf W_{i,t}$. Each $\textbf W_{i,t}$ is characterized with one of five possible windowing type functions, $t$, on the $i^{th}$ window segment, where $\forall i \in \overline{1,C+1}$ . The $i^{th}$ window segment is associated to $\textbf S$'s spectral data by $[S_{1+(i-1) \cdot L_{w}},S_{i \cdot L_{w}(i-1)+L_w}]$.
\begin{equation}
\textbf W_{i,t}=[ W_{1}, \hspace{5 mm}  W_{2}, \hspace{2 mm}  \cdots \hspace{2 mm}  W_{L_w}]^T,
\end{equation}
where, $ W_j\in[0,1], \forall j \in \overline{1,L_w}, \forall t \in \overline{1,5},  \forall i \in \overline{1,C}$. Equation~\ref{eq:specialcase}, handles the residual data of $S$, since we only designed $C$ windows.
\begin{equation} \label{eq:specialcase}
\textbf W_{(C+1),t}=[ W_{1}, \hspace{5 mm}  W_{2}, \hspace{2 mm}  \cdots \hspace{2 mm}  W_{N-(L_w \cdot C)}]^T,
\end{equation}
where, $ W_j\in[0,1], \forall j \in \overline{1,L_w}, \forall t \in \overline{1,5},  \forall i \in \overline{1,C}$.
The type of windowing function that is implemented for each $i$ window segment is determined by recursively examining each case in numerical order until the specific conditions of a case is in accordance of the criteria. The windowing functions and conditioned criteria are defined by the following five cases, where for all cases $\alpha \in[0,1]$, $\gamma=(L_w -1)$, and $\beta=(1-\frac{\alpha}{2})$.
\\
\\
\textbf{\emph{Windowing Case 1: Hanning Window}}
\begin{equation}
\textbf W_{i,1}(j)=.5(1-\cos(2\pi(\frac{j}{L_w})))
\end{equation}
$\mbox{\textit{if and only if}}\hspace{1 mm}  \exists i \in \textbf{I}_p$\\
\\
\textbf{\emph{Windowing Case 2: Tukey Window}}
\begin{equation}
\textbf W_{i,2}(j)= \left\{ \begin{array}{ll}
\frac{1+\cos{(\pi ( \frac{2(j-1)}{\alpha(\gamma)} )-1 )}}{2},          &\mbox{for $1 \leqslant j \leqslant \frac{\alpha (\gamma)}{2} $}\\
1,                               &\mbox{for $ \frac{\alpha (\gamma)}{2}\leqslant j \leqslant (\gamma)(\beta)$}  \\
 \frac{1+\cos{(\pi ( \frac{2(j-1)}{\alpha(\gamma)})-\frac{2}{\alpha}+1 )}}{2}, &\mbox{for $ (\gamma)(\beta) \leqslant j \leqslant (\gamma)$}\\
       \end{array} \right.
\end{equation}
$\mbox{\textit{if and only if}}\hspace{1 mm}  \exists i \not \in \textbf{I}_p \wedge \exists (i-1) \in \textbf{I}_p \wedge \exists (i+1)  \in \textbf{I}_p $\\
\\
\textbf{\emph{Windowing Case 3: Modified Left Tukey Window}}
\begin{equation}
\textbf W_{i,3}(j)= \left\{ \begin{array}{ll}
\frac{1+\cos{(\pi ( \frac{2(j-1)}{\alpha(\gamma)} )-1 )}}{2},          &\mbox{for $1 \leqslant j \leqslant \frac{\alpha (\gamma)}{2} $}\\
1,                               &\mbox{for $ \frac{\alpha (\gamma)}{2}\leqslant j \leqslant (\gamma) $}  \\
       \end{array} \right.
\end{equation}
$\mbox{\textit{if and only if}}\hspace{1 mm}  \exists i \not \in \textbf{I}_p \wedge \exists (i-1)  \in \textbf{I}_p \wedge \exists (i+1) \not \in \textbf{I}_p $\\
\\
\textbf{\emph{Windowing Case 4: Modified Right Tukey Window}}
\begin{equation}
\textbf W_{i,4}(j)= \left\{ \begin{array}{ll}
1,                               &\mbox{for $ 1 \leqslant j \leqslant (\gamma)(\beta)$}  \\
 \frac{1+cos{(\pi ( \frac{2(j-1)}{\alpha(\gamma)})-\frac{2}{\alpha}+1 )}}{2}, &\mbox{for $ (\gamma)(\beta) \leqslant j \leqslant(\gamma)$}\\
       \end{array} \right.
\end{equation}
$\mbox{\textit{if and only if}}\hspace{1 mm}  \exists i \not \in \textbf{I}_p \wedge \not \exists (i-1) \in \textbf{I}_p \wedge \exists (i+1)  \in \textbf{I}_p$
\\
\\
\textbf{\emph{Windowing Case 5: Null Variance Window}}
\begin{equation}
\textbf W_{i,5}(j)= 1, \hspace{12 mm} \mbox{for $\forall j$}  \\
\end{equation}
$\mbox{\textit{if and only if}}\hspace{1 mm}  \exists i \not \in \textbf{I}_p \wedge \exists (i-1) \not \in \textbf{I}_p \wedge \exists (i+1)  \not \in \textbf{I}_p$
\\
\\
Each individual $\textbf W_{i,t}$ window vector that is designed will be cascaded in numerical order to construct, $\textbf J_F$, the foundation for creating our dilation compression vector to modify  vector $\textbf  S$. Hence, $\textbf  J_F$ is defined as:
\begin{equation} \label{eq:WindowingFoundation}
\textbf J_F=[ \textbf W_{1,t}, \hspace{2 mm}  \textbf W_{2,t}, \hspace{5 mm}   \cdots \hspace{2 mm} \textbf W_{C,t}, \hspace{2 mm} \textbf W_{(C+1),5}]^T,
\end{equation}
where $ \textbf W_{i,t}\in[0,A], \forall i \in \overline{1,C},  \forall t \in \overline{1,5},$  and $\textbf J\in[0,A], \forall i \in \overline{1,(C+1)}$. However, note we only design $C$ windows, the (C+1) window will always default to case $5$ to handle the residual data of $\textbf S$ thats smaller than the specified window size, $L_w$.

\subsection{Basic Design for Dilation Noise}
Once our foundation vector is set, we modify, $W_{i,t}$. We demonstrate a few different methods to modify $W_{i,t}$ to create the most ideal synthetic noise representation starting from the most basic. As well, we review the down falls in order to improve upon each method. We first exemplify synthetic noise to the data by simply dilating the window locations where the peaks occur. This dilation is scaled by a constant $\eta$. Therefore, we revamp $W_{1,t}$, to be scaled by $\eta$ and force other Tukey windows $W_{2-4,t}$ to one:
\begin{gather}
\hat{ \textbf W}_{\hat{D}_{1,t}}= (\eta) \cdot \textbf W_{1,t} + U_{L_w}, \label{eq:W1Dilation} \\
\hat{ \textbf W}_{\hat{D}_{2-5,t}}=(0) \cdot \textbf W_{2-5,t} + U_{L_w}, \\
\textbf J_{\hat{D}}=[\hat{ \textbf W}_{\hat{D}_{1,t}},\hspace{2mm}\hat{\textbf W}_{\hat{D}_{2,t}},\hspace{5mm}\cdots\hspace{2mm}\hat{\textbf W}_{\hat{D}_{C,t}},\hspace{2mm}\textbf W_{(C+1),5}]^T, \\
\textbf S_{S+N}=\textbf J_{\hat{D}} \hspace{1mm} diag(\textbf S)
\end{gather}

\begin{figure}[H]
\centering
\includegraphics [width=.9\linewidth]{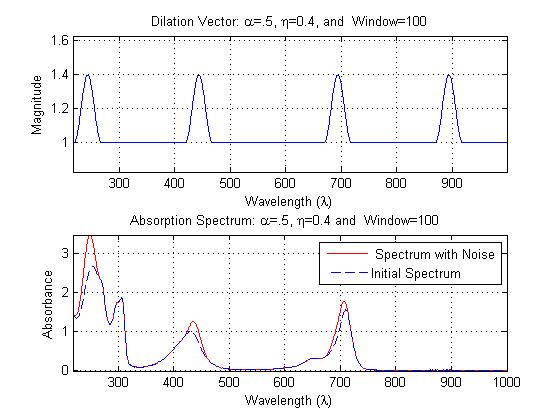}
\caption{Absorbance Spectrum with Dilation Noise.}\label{fig:DilNoise}
\end{figure}
We can see in figure~\ref{fig:DilNoise}, that by implementing this method we are only given the option of dilating the spectrum at a constant $\eta$. In addition, its worth noting due to the fixed windowing segments, the alignment of the peaks are not centered directly over the windowed segment causing shifting of the peak wavelength. In some scenarios, this maybe considered ideal for further perturbations of the signal, where the shifting of the peak wavelength is a function of $\eta$.

\subsection{Basic Design for Compression Noise}

Given that we can dilate the signal, in order for us to compress the signal, a compression vector is designed as such by equation~\ref{eq:ComNoiseA}, \ref{eq:ComNoiseB}, and \ref{eq:ComNoiseC}, which is pictorially represented in Figure~\ref{fig:ComNoise}:
\begin{gather}
\hat{ \textbf W}_{\hat{C}_{i,t}}=\eta \cdot \textbf W_{i,t}+(U_{L_w}-\eta U_{L_w}) \label{eq:ComNoiseA}\\
\textbf J_{\hat{C}}=[ \hat{ \textbf W}_{\hat{C}_{1,t}}, \hspace{2mm}  \hat{\textbf W}_{\hat{C}_{2,t}}, \hspace{5mm}   \cdots \hspace{2mm} \hat{\textbf W}_{\hat{C}_{C,t}}, \hspace{2mm} \textbf W_{(C+1),5}]^T, \label{eq:ComNoiseB} \\
\textbf S_{S+N}=\textbf J_{\hat{C}}  \hspace{1mm} \textit{diag}(\textbf S)  \label{eq:ComNoiseC}
\end{gather}
where $D_k\in[0,A], \forall k \in \overline{1,N}$
\begin{figure}[h]
\centering
\includegraphics [width=4.0in]{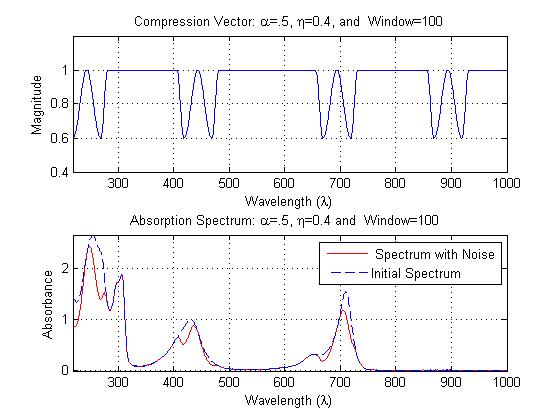}  
\caption{Absorbance Spectrum with Compression Noise.}\label{fig:ComNoise}
\end{figure}

\subsection{An Axiomatic Approach for Designing Appropriate Noise}

In the previous sections, we discussed the foundation of the windowing and the functions that were integral for dilation and compression of the signal by a factor of $\eta$.  It's only sensible to anticipate the above methods as a combination of random dilation and compression for a more convincing simulated noise. Considering these two functions as a whole, comprised of different adjacent windowing techniques, it is careless to assume that $\eta$'s effect on the signal is equivalent. A pragmatic approach of declaring guidelines for the noise algorithm must be established to meet the appropriate standards to have unbiased detection simulations when noise is introduced.  In order to fortify this concept of why guidelines are required and to exam what guidelines that need to be put in place, we introduce an algorithm that functionally fails as a dilating compressing algorithm.

\subsubsection{Failure for Proper Dilation and Compression}

\begin{gather}
 \hat{ \textbf W}_{i,1}= \eta X \textbf W_{i,1}+(1- \frac{\eta}{2})U_{L_w} \label{eq:DilationFailure1},\\
 \hat{ \textbf W}_{i,2-4}= (\eta) \textbf W_{i,2-4}+(1- \frac{\eta}{2})U_{L_w})  \label{eq:DilationFailure2},\\
  \hat{\textbf W}_{i,5}= \textbf W_{i,5},  \label{eq:DilationFailure2}, \\
 \textbf J_{\hat{CD}}=[\hat{ \textbf W}_{\hat{D}_{1,t}},\hspace{2mm}\hat{\textbf W}_{\hat{D}_{2,t}},\hspace{5mm}\cdots\hspace{2mm}\hat{\textbf W}_{\hat{D}_{C,t}},\hspace{2mm}\textbf W_{(C+1),5}]^T, \\
\textbf S_{S+N}=\textbf J_{\hat{CD}}\hspace{1mm} \textit{diag}(\textbf S)
\end{gather}
where X is a random variable uniformly distributed $X \in[0,1]$
\begin{figure}[h]
\centering
\includegraphics[width=.9\linewidth]{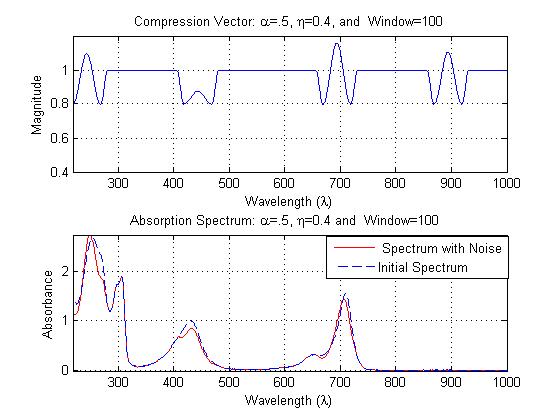}
\caption{Absorbance Spectrum with Noise Failing to Dilate.}\label{fig:ComNoise}
\end{figure}
Visually examining the noise vector, we can note that the hanning windows exceeds the peak magnitude of one.  We may assume the data vector will undergo dilation and compression, but this is deceiving. When the filter is applied to the data vector the data is not properly dilated as expected. It begins to become apparent, graphically in Figure~\ref{fig:ComNoise} that, the  magnitude of dilation on the sample may not be comparable to the compression. However, the improper dilation may be attributed to that particular sample, therefore simulations are done in the preceding section to examine how the intensity changes are distributed across numerous samples.

\subsubsection{Noise Design Guidelines}

The previous example shows the possible cause of how the noise is distributed over the vector. When examined over numerous trials, it is possible that it may not be ideal for testing our detection algorithm in the later chapters. Hence, we set guidelines of what our ideal noise distribution should be in order to achieve an appropriate testing scenario for our detection algorithm. The guidelines are verified by visual inspection from simulations that produce the noise distribution.

\begin{figure}[H]
\centering
\includegraphics [width=.9\linewidth]{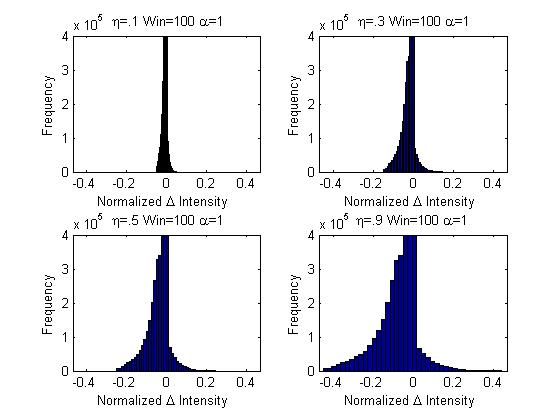}
\caption{Histogram of the Change in Intensity of Corresponding Elements.}\label{fig:NoiseDistFail}
\end{figure}

\textbf{{\emph{ Synthetic Noise Distribution Ideal Guidelines}}}
\begin{enumerate}
  \item $\eta$ will be random for each designed window.
  \item Within a single noise vector, Dilation and Compression can occur at various locations.
  \item As $\eta$ increases, the variance of the distribution will increase.
  \item The noise distribution when evaluating the magnitude between corresponding elements should be symmetric around zero.
  \item The distribution of intensity loss and gain for a entire vector should be symmetric around zero.
\end{enumerate}
These guidelines will assist in the development of a pragmatic noise that will better challenge the subsequent detection process. The simulation that is imposed to validate if the noise distribution meets the guideline criteria is created with a database size of eleven Chemicals yielding $2^{Chem}-1=2047$ chemical combinations , to process $6000$ randomly picked chemicals. The correlated noise algorithm applied the following noise equations \ref{eq:DilationFailure1} ,\ref{eq:DilationFailure1}, and \ref{eq:DilationFailure1}. A normalization for element intensity in equation \ref{eq:NormMagnitude} was imposed, since the absorbance can vary so greatly, $N_{EI}$.
\begin{equation}
\textbf N_{EI}=\frac {\textbf{S}} {\textsl{max}( \textbf S )} \label{eq:NormMagnitude}
\end{equation}
In order to create a robust synthetic noise model, we need to account for dilation and compression equally. We are able to note the displacement of the intensity to the corresponding elements with a histogram representation of $\Delta \textbf I_E$, defined as \ref{eq:DifferenceElement}. The parameter
$\eta$ is varied over four simulation to evaluate the effects of the variance on the distribution, as shown in Figure~\ref{fig:NoiseDistFail}, identifying the asymmetry in the noise distribution.  This current noise model does not meet the guidelines stated and is therefore non-ideal. The model provided would provide questionable results in the detection algorithm, where it is ambiguous if the algorithm failed or if it was a slight change in $\eta$ causes a large shift in the attenuation of the original signal.
\begin{equation}
 \Delta \textbf I_E= \textbf J_{ \hat{CD}} \hspace{1mm} \textit{diag}(\textbf N_{EI})- \textbf N_{EI} \label{eq:DifferenceElement}
\end{equation}

\begin{figure}[H]
\centering
\includegraphics [width=.9\linewidth]{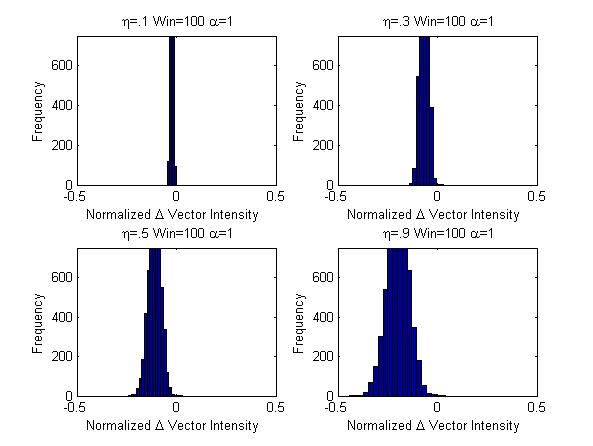}
\caption{Histogram of the Change in Intensity of Entire Vector.}\label{fig:NoiseEnergyFail}
\end{figure}

 Figure~\ref{fig:NoiseEnergyFail} examines the change in intensity over the entire vector, which solidifies the need for the stated guidelines of $(4)$ and $(5)$. We can can see that from the distribution that a shifting is occurring as $\eta$ increases, preventing any type of dilation to occur within the spectrum. The change of intensity over the entire vector, $\textbf N_{VI}$ , of the normalized spectrum, $\textbf N_{VI}$, was calculated by equations \ref{eq:NormEngery} and \ref{eq:DifferenceEngery}, to achieve the histogram Figure~\ref{fig:NoiseEnergyFail}.

\begin{gather}
\textbf N_{VI}=\frac { \textbf{S}} {\sum_{k=1}^{N}( \textbf S )} \label{eq:NormEngery}\\
\Delta \textbf I_V= \sum_{k=1}^{N}(\textbf J_{\hat{CD}} \hspace{1mm} \textit{diag}(\textbf N_{VI}) - \textbf N_{VI}) \label{eq:DifferenceEngery}
\end{gather}

\begin{figure}[H]
\centering
\includegraphics [width=.9\linewidth]{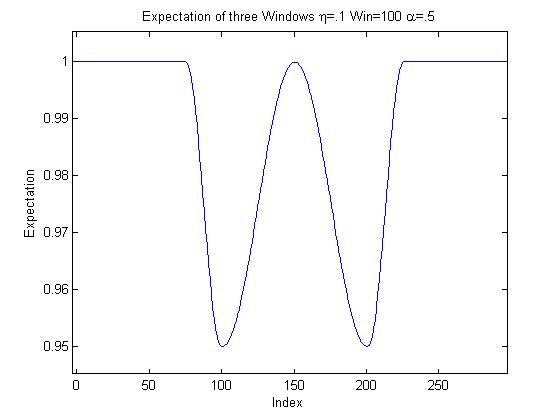}
\caption{Windowing Segment Examining Expectation.}\label{fig:ExpectationWindow}
\end{figure}

The cause can be further explained by examining the expectation of the designed noise vector $W_{i,1}$ elements' $j$ and the adjacent windowing vectors.
\begin{align*}
E[\hat{ \textbf W}_{{i,1}_j}] &= E[\eta \cdot X \cdot W_{{i,1}_j} + U_j- \frac{\eta}{2} U_j]  \\
E[\hat{ \textbf W}_{{i,1}_j}] &= \eta \cdot W_{{i,1}_j} E[X] + E[U_j]- E[\frac{\eta}{2} U_j]  \\
E[\hat{ \textbf W}_{{i,1}_j}] &= \eta \cdot W_{{i,1}_j} E[X] + E[1]- E[\frac{\eta}{2}]   \\
E[\hat{ \textbf W}_{{i,1}_j}] &= \eta \cdot W_{{i,1}_j} \frac{1}{2} + 1- \frac{\eta}{2}
\end{align*}
The adjacent windowing elements are constants and do not need to be evaluated ($W_{i,3}$ $W_{i,4}$), but should be examined pictorially with $W_{i,1}$ in Figure~\ref{fig:ExpectationWindow}.

We can now note, pictorially from Figure~\ref{fig:ExpectationWindow}, that none of the elements expectations surpass one, ultimately causing a summed loss of intensity or compression of the signal. We can also note that anytime these adjacent windows occur ($W_{i,2}$ $W_{i,3}$ $W_{i,4}$), we will always encounter a further induced intensity loss. This intensity lose is not equally compensated by dilation. Furthermore, even when dilation occurs within the function \ref{eq:DilationFailure1} it still compresses the width of the peak, even though it's an effective method for dilation. This deficiency can be attributed to the innate way we window, since the peak of the signal can range anywhere with the window length $L_w$. However, we should not see this as a shortcoming since it causes further realist noise by shifting the spectrum by a function of $\eta$.

\subsection{Approximating an Advantageous Synthetic Noise}

We propose the following combined dilation and compression noise functions to fulfill the following guidelines, based on the windowing foundation function \ref{eq:WindowingFoundation}. In order to satisfy these requirements, constraints are imposed to the functions. However, we maintain the functions robustness to dynamically be altered for compression and dilation.

\begin{figure}[ht]
\centering
\includegraphics[width=.9\linewidth]{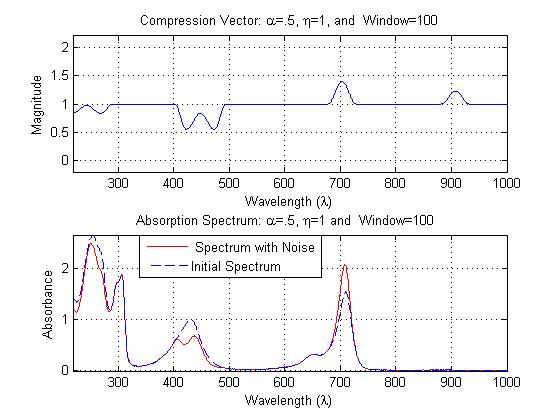}
\caption{Simulated Optimized Noise.}\label{fig:Noiseopt}
\end{figure}

\begin{figure}[ht!]
        \centering
        \subfloat[Element Intensity]
        {\label{fig:NoiseCorrectionElement}\includegraphics[width=.9\linewidth]{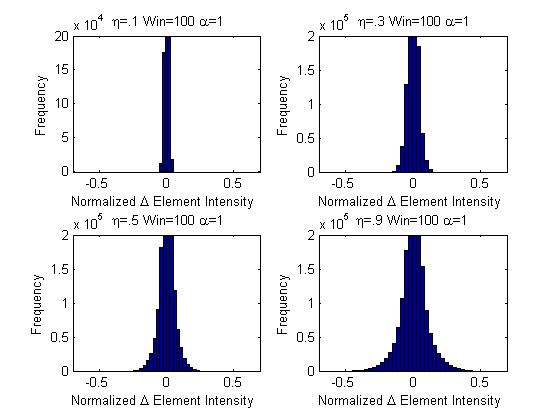}}\qquad
        \subfloat[Vector Intensity]{\label{fig:NoiseCorrectionvector}\includegraphics[width=.9\linewidth]{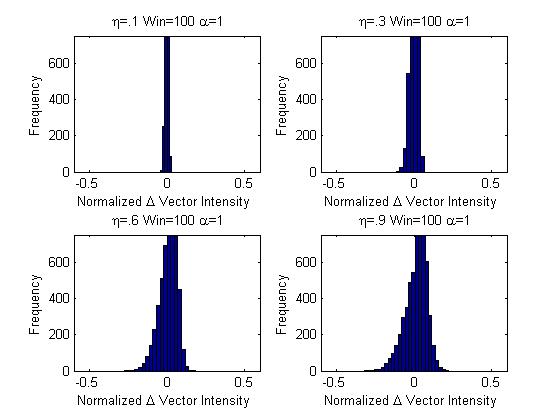}}\qquad
        \caption{Histograms of Intensity.}
        \label{fig:multipart-figure}
\end{figure}

\subsubsection{Defining a Robust function}

This is done by using a binomial distribution to create an indictor function $\textbf I$ to determine what $\textbf W_{i,1}$ function to implement, compression  $\hat{ \textbf W}_{C_{i,1}}$ or dilation $\hat{ \textbf W}_{D_{i,1}}$, where $p=.5$ , in equation \ref{eq:Binomial}. This allows further manipulation of the function for us to engage the guidelines criteria more stringently by enabling two functions to competent against each other to achieve an approximate zero mean distribution.

\begin{equation} \label{eq:Binomial}
\textmd{I}= \left\{ \begin{array}{ll}
1,           &\mbox{with probablity p}  \\
0,           &\mbox{with probablity (1-p)}\\
       \end{array} \right.
\end{equation}
The adjacent window's elements are dependent on the design of $\hat{ \textbf W_{i,1}}$. We designed $\hat{ \textbf W_{i,1}}$ from the fundamentals of \ref{eq:W1Dilation} and \ref{eq:ComNoiseA} for the compression and dilation vectors. Thus, the following equations~\ref{eq:OptCompressor}, \ref{eq:OptDilator} were developed for the $ \textbf W_{i,1}$ windowing case.
\begin{gather}
 \hat{ \textbf W}_{B_{i,1}}= \hat{ \textbf W}_{C_{i,1}} \textmd I +  \hat{ \textbf W}_{D_{i,1}} (\textmd I -1)      \label{eq:Binomial}\\
 \hat{ \textbf W}_{C_{i,1}}= (A_{c1}X_2 \eta) \textbf W_{i,1}+ \textbf U_{L_w}(1-A_{c2}\eta+A_{c3}\eta X_3)  \label{eq:OptCompressor}\\
 \hat{ \textbf W}_{D_{i,1}}= (A_{d1}X_1 \eta) \textbf W_{i,1}+ \textbf U_{L_w}  \label{eq:OptDilator}
\end{gather}
where $\eta \in[0,1],X_i$ is a uniform random variable $\in[0,1], \forall i \in \overline{1,3}$, and $A_xi$ are coefficients to control the functions. In order to avoid discontinuities between the adjacent windows, due to biasing from each window from the random variable $X_3$, the following adjustments were made to the following cases:
\begin{gather}
 \hat{ \textbf W}_{i,2}= \textbf W_{i,2} (1- \hat{ \textbf W}_{i+1,1}(1)) +    \hat{ \textbf W}_{i+1,1}(1), \hspace{8mm} \mbox{for $ (\gamma)(\beta) \leqslant j \leqslant (\gamma)$} \\
 \hat{ \textbf W}_{i,2}= \textbf W_{i,2} (1- \hat{ \textbf W}_{i-1,1}(L_W)) +  \hat{ \textbf W}_{i-1,1}(L_W), \hspace{8mm} \mbox{for $1 \leqslant j \leqslant \frac{\alpha (\gamma)}{2} $}\\
 \hat{ \textbf W}_{i,3}= \textbf W_{i,3} (1- \hat{ \textbf W}_{i+1,1}(1)) +    \hat{ \textbf W}_{i+1,1}(1),\\
 \hat{ \textbf W}_{i,4}= \textbf W_{i,4} (1- \hat{ \textbf W}_{i-1,1}(L_W)) +  \hat{ \textbf W}_{i-1,1}(L_W),
\end{gather}

\subsubsection{Constraint A:}

Constraint A is to bound the expected dilation and compression functions to be equivalent. This will assist one functions maximum from overpowering the other and maintain a equivalent magnitude changes.

\begin{align*}
1-E[\min_{ j \in \overline{1,L_W}} (\hat{W}_{C_{i,1}j})]& = E[ \max_{ j \in \overline{1,L_W}}(\hat{W}_{D_{i,1}})]-1\\
1-E[1-A_{c2}\eta+A_{c3}\eta X_3] & =E[ A_{d1}X_1 \eta+ 1 ]-1\\
1-E[1]+E[A_{c2}\eta]-E[A_{c3}\eta X_3]&=E[ A_{d1}X_1 \eta]+ E[1]-1\\
E[A_{c2}\eta]-E[A_{c3}\eta X_3]&=E[ A_{d1}X_1 \eta]\\
\eta A_{c2}-A_{c3}\eta E[X_3]&=A_{d1}\eta E[X_1]\\
A_{c2}-A_{c3}E[X_3]&=A_{d1}E[X_1]\\
A_{c2}-\frac{1}{2}A_{c3}&=\frac{1}{2}A_{d1}\\
2A_{c2}-A_{c3}&=A_{d1}\\
\end{align*}

\subsubsection{Constraint B:}

Constraint B is to bound the variance of the dilation and compression functions to be equivalent. This will prevent the dilation and compression functions from causing an asymmetry within the noise distribution.

\begin{align*}
Var[\min_{ j \in \overline{1,L_W}} (\hat{W}_{C_{i,1}j})]& = Var[ \max_{ j \in \overline{1,L_W}}(\hat{ W}_{D_{i,1}})]\\
Var[1-A_{c2}\eta+A_{c3}\eta X_3] & =Var[ A_{d1}X_1 \eta+ 1 ]\\
Var[A_{c3}\eta X_3] & =Var[ A_{d1}X_1 ]\\
A_{c3}^2\eta^2 Var[X_3]& =A_{d1}^2 \eta^2 Var[X_1]\\
A_{c3}^2 & =A_{d1}^2
\end{align*}

\subsubsection{Constraint C:}

Constraint C is to bound the compression function's maximum element in that vector to never surpass the maximum value of the dilation function's maximum element.
\begin{align*}
\min_{ j \in \overline{1,L_W}} (\hat{W}_{C_{i,1}j})& \leq \max_{ j \in \overline{1,L_W}}(\hat{ W}_{D_{i,1}})\\
A_{c1}\eta + 1-A_{c2}\eta+A_{c3}\eta X_3 & \leq  A_{d1}X_1 \eta+ 1 \\
A_{c1}\eta -A_{c2}\eta + A_{c3}\eta X_3 & \leq  A_{d1}X_1 \eta\\
A_{c1} - A_{c2} + A_{c3} & \leq  A_{d1}
\end{align*}

\subsubsection{Constraint D:}

Recalling Figure~\ref{fig:ExpectationWindow}, based on the topology of design there will always be a greater intensity loss with the compression function is implemented. To achieve an approximate zero mean intensity loss for the vector, stated by the fifth guideline, we minimize the expected intensity loss of the vector when compared in a random process of dilation and compression by exploiting the binomial equation. Using the binomial equation to our advantage, we can design the $\hat{ \textbf W}_{C_{i,1}}$ and $\hat{ \textbf W}_{D_{i,1}}$ functions accordingly having them achieve an approximate averaged expected intensity value over all elements, \ref{eq:AverageExpect}. In order to account for this average expectation, we account for the effects of the adjacent windows as well for a proper approximation. In the dilations case we used $ \textbf W_D=[\textbf W_{i,5}, \hat{ \textbf W}_{i,1},  \textbf W_{i,5}]$  and for the compression case we used $\textbf W_C=[\hat{ \textbf W}_{i,3}, \hat{ \textbf W}_{i,1},  \hat{ \textbf W}_{i,4}]$. Therefore, base on the design of $\hat{ \textbf W}_{i,1}$ for the compression and dilation we can manipulate our indictors function's $p$ value to further optimize the expected intensity to have an approximate equivalent intensity deviation of compression and dilation.

\begin{equation} \label{eq:AverageExpect}
E_{win}=p \frac{1}{3*L_w}(\sum_{j=1}^{3 \cdot Lw}E[W_C] )+ (1-p)\frac{1}{3*L_w}(\sum_{j=1}^{3 \cdot Lw} E[W_D]) \approx 1
\end{equation}
The following approximation was done with $\alpha=1,\eta=.1,A_{c1}=A_{c2}=A_{c3}=A_{d1}=.5$ and $L_w=100$, where $E_{win}=0.9918$ with a $p=.33$, implementing this optimize noise at these parameters we manifest others noise vectors at various different $\eta$'s assuming the changes are minute. If desired you can optimize the $p$ for different $eta$s. In Figure~\ref{fig:Noiseopt}, we used the aforementioned parameters to designed noise at $\eta=1$, where simulations of the distributions of element and vector intensity is shown in \ref{fig:multipart-figure}.


\section{Conclusion}
The synthetic noise generation was designed to create to mimic spectrum's changes caused by user error, stray light, wavelength accuracy, and self absorption. As you can note, that additive white Gaussian noise simply can not use to mimic such erroneous changes within the spectroscopy signal. The noise that was created is correlated to the various peaks within the spectrum and then constrained by the percentage of error by the parameter $\nu$. More completed work that highlights the full creation of these spectroscopy data sets and probably detection applications are available~\cite{NapoliThesis,NapoliFusion15}.

\bibliographystyle{unsrtnat}
\bibliography{references}

\begin{thebibliography}{22}
\providecommand{\natexlab}[1]{#1}
\providecommand{\url}[1]{\texttt{#1}}
\expandafter\ifx\csname urlstyle\endcsname\relax
  \providecommand{\doi}[1]{doi: #1}\else
  \providecommand{\doi}{doi: \begingroup \urlstyle{rm}\Url}\fi

\bibitem[Du et~al.(1998)Du, Fuh, Li, Corkan, and Lindsey]{Du}
Hai Du, Ru-Chun~Amy Fuh, Junzhong Li, Andrew Corkan, and Jonathan~S. Lindsey.
\newblock Photochemcad. a computer-aided design and research tool in
  photochemistry and photobiology.
\newblock \emph{Photochem.Photobiology}, 68:\penalty0 141--142, 1998.

\bibitem[Dixon et~al.(2005)Dixon, Taniguchi, and Lindsey]{Dixon}
J.M. Dixon, M.~Taniguchi, and J.S. Lindsey.
\newblock Photochemcad two. a refined program with accompanying spectral
  databases for photochemical calculations.
\newblock \emph{Photochem.Photobi}, 81:\penalty0 212--213, 2005.

\bibitem[Taniguchi et~al.(2002)Taniguchi, Kim, D.~Ra, Kirmaier, Hindin, Diers,
  Prathapan, Bocian, Holten, and Lindsey]{Taniguchi}
M.~Taniguchi, H.-J. Kim, J.~K.~Schwartz D.~Ra, C.~Kirmaier, E.~Hindin, J.~R.
  Diers, S.~Prathapan, D.~F. Bocian, D.~Holten, and J.~S. Lindsey.
\newblock Synthesis and electronic properties of regioisomerically pure
  oxochlorins.
\newblock \emph{J. Organic Chemistry}, 67:\penalty0 7329--7342, 2002.

\bibitem[Strachan et~al.(2000)Strachan, OShea, Balasubramanian, and
  Lindsey]{Strachana}
J.P. Strachan, D.~F. OShea, T.~Balasubramanian, and J.~S. Lindsey.
\newblock Rational synthesis of meso-substituted chlorin building blocks.
\newblock \emph{J. Organic Chemistry}, 65:\penalty0 3160--3172, 2000.

\bibitem[Sazanovich et~al.(2004)Sazanovich, Kirmaier, Hindin, Yu, Bocian,
  Lindsey, and Holten]{Sazanovich}
I.~V. Sazanovich, C.~Kirmaier, E.~Hindin, L.~Yu, D.~Bocian, J.~S. Lindsey, and
  D.~Holten.
\newblock Structural control of the excited-state dynamics of
  bis(dipyrrinato)zinc complexes: self-assembling chromophores for
  light-harvesting architecture.
\newblock \emph{J. Am. Chem. Soc.}, 126:\penalty0 2664--2665, 2004.

\bibitem[Zass et~al.(1990)Zass, Isenring, Etter, and Eschenmoser]{Zass}
E.~Zass, H.~P. Isenring, R.~Etter, and A.~Eschenmoser.
\newblock Der einbau van magnesium in liganden der chlorophyll-reihe mit
  (2,6-di-t-butyl-4-methylphenoxy)magnesiumjodid.
\newblock \emph{Helv. Chim. Acta}, 63:\penalty0 1048--1067, 1990.

\bibitem[Yang et~al.(1999)Yang, Seth, Strachan, Gentemann, Kim, Holten,
  Lindsey, and Bocian]{Yang}
S.~I. Yang, J.~Seth, J.-P. Strachan, S.~Gentemann, D.~Kim, D.~Holten, J.~S.
  Lindsey, and D.~F. Bocian.
\newblock Ground and excited state electronic properties of halogenated
  tetraarylporphyrins: Tuning the building blocks for porphyrin-based
  nanostructures.
\newblock \emph{J. Porphyrins Phthalocyanines}, 3:\penalty0 117--147, 1999.

\bibitem[Lindsey and Woodford(1995)]{Lindsey}
J.~S. Lindsey and J.~N. Woodford.
\newblock A simple method for preparing magnesium porphyrins.
\newblock \emph{Inorg. Chem.}, 34:\penalty0 1063--1069, 1995.

\bibitem[Miller and Dorough(1952)]{Miller}
J.~R. Miller and G.~D. Dorough.
\newblock Pyridinate complexes of some metallo-derivatives of
  tetraphenylporphine and tetraphenylchlorin.
\newblock \emph{J. Am. Chem. Soc.}, 74:\penalty0 3977--3981, 1952.

\bibitem[Strachan et~al.(1997)Strachan, Gentemann, Seth, Kalsbeck, Lindsey,
  Holten, and Bocian]{Strachanb}
J.~P. Strachan, S.~Gentemann, J.~Seth, W.~A. Kalsbeck, J.~S. Lindsey,
  D.~Holten, and D.~F. Bocian.
\newblock Effects of orbital ordering on electronic communication in
  multiporphyrin arrays.
\newblock \emph{J. Am. Chem. Soc.}, 119:\penalty0 11191--11201, 1997.

\bibitem[Prathapan et~al.(2001)Prathapan, Yang, Seth, Miller, Bocian, Holten,
  and Lindsey]{Prathapan}
S.~Prathapan, I.~Yang, J.~Seth, M.~A. Miller, D.~F. Bocian, D.~Holten, and
  J.~S. Lindsey.
\newblock Synthesis and excited-state photodynamics of perylene-porphyrin
  dyads. 1. parallel energy and charge transfer via a diphenylethyne linker.
\newblock \emph{Journal of Physical Chemistry}, B 105:\penalty0 8237--8248,
  2001.

\bibitem[Tomizaki et~al.(2002)Tomizaki, Loewe, Kirmaier, Schwartz, Retsek,
  Bocian, Holten, and Lindsey]{Tomizaki}
K.~Tomizaki, R.~S. Loewe, C.~Kirmaier, J.~K. Schwartz, J.~L. Retsek, D.~F.
  Bocian, D.~Holten, and J.~S. Lindsey.
\newblock Synthesis and photophysical properties of light-harvesting arrays
  comprised of a porphyrin bearing multiple perylene-monoimide accessory
  pigments.
\newblock \emph{Journal of Organic Chemistry}, 67:\penalty0 6519--6534, 2002.

\bibitem[Reusch(2012)]{Reusch}
W.~Reusch.
\newblock Michigan state university: Visible and ultraviolet spectroscopy.
\newblock July 2012.
\newblock URL
  \url{http://www2.chemistry.msu.edu/faculty/reusch/VirtTxtJml/intro1.htm}.

\bibitem[Szabo(2000)]{Szabo}
Arthur~G. Szabo.
\newblock \emph{Spectrophotometry and Spectrofluorimety}, chapter Fluroscence
  principle and measurement, page~40.
\newblock 2000.

\bibitem[Leon-Garcia(2008)]{Garcia}
A.~Leon-Garcia.
\newblock \emph{Probability,Statistics and Random Processes for Electrical
  Engineering}.
\newblock 2008.

\bibitem[Napoli and Barnes(2016)]{Napoli_DS_EKG}
Nicholas~Joseph Napoli and Laura~E Barnes.
\newblock {A Dempster-Shafer Approach for Corrupted Electrocardiograms
  Signals}.
\newblock \emph{Twenty-Ninth International Florida Artificial Intelligence
  Research Society Conference}, 2016.

\bibitem[Napoli et~al.(2016)Napoli, Leach, Barnes, and Weimer]{NapoliMap}
Nicholas~J. Napoli, Kevin Leach, Laura~E. Barnes, and Westley Weimer.
\newblock A mapreduce framework to improve template matching uncertainty.
\newblock \emph{Big Data and Smart Computing}, 2016.

\bibitem[Lakowicz(2006)]{Lakowicz}
J.~R. Lakowicz.
\newblock \emph{Principles of Fluorescence Spectroscopy}.
\newblock 2006.

\bibitem[Barnes et~al.(1989)Barnes, Dhanoa, and Lister]{Barnes}
R.J. Barnes, M.S. Dhanoa, and S.J. Lister.
\newblock Standard normal variate transformation and de-trending of
  near-infared diffuse reflectance spectra.
\newblock \emph{Applied Spectroscopy}, 43:\penalty0 772--777, 1989.

\bibitem[Allen(2012)]{Allen}
M.W. Allen.
\newblock Stray light-measurement and effect on performance in uv-visible
  spectrophotometry.
\newblock TechnicalNote 51170, Thermo Fisher Scientific, Madison, WI, USA, July
  2012.

\bibitem[Napoli(2014)]{NapoliThesis}
Nicholas~J. Napoli.
\newblock The detection of analytes using spectroscopy: A dempster-shafer
  approach, 2014.
\newblock URL \url{http://scholarlyrepository.miami.edu/oa_theses/517}.

\bibitem[Napoli et~al.(2015)Napoli, Barnes, and Premaratne]{NapoliFusion15}
Nicholas~J. Napoli, Laura~E. Barnes, and Kamal Premaratne.
\newblock Correlation coefficient based template matching: Accounting for
  uncertainty in selecting the winner.
\newblock \emph{Fusion}, pages 311--318, 2015.

\end{thebibliography}

\end{document}